\documentclass[useAMS,usegraphicx,usenatbib]{mn2e}

\title[Be 23, Be 31, and King 8]{Three new bricks in the
  wall:\\ Berkeley 23, Berkeley 31, and King
  8\thanks{Based on observations collected at the LBT
    and in part at the TNG. The Large Binocular Telescope (LBT) is an
    international collaboration among institutions in the United
    States, Italy and Germany. LBT Corporation partners are: The
    University of Arizona on behalf of the Arizona university system;
    Istituto Nazionale di Astrofisica, Italy; LBT
    Beteiligungsgesellschaft, Germany, representing the Max-Planck
    Society, the Astrophysical Institute Potsdam, and Heidelberg
    University; The Ohio State University, and The Research
    Corporation, on behalf of The University of Notre Dame, University
    of Minnesota and University of Virginia. The Italian Telescopio
    Nazionale Galileo (TNG) is operated on the island of La Palma by
    the Fundaci\'on Galileo Galilei of the INAF (Istituto Nazionale di
    Astrofisica) at the Spanish Observatorio del Roque de los
    Muchachos of the Instituto de Astrofisica de Canarias} }

   \author[Michele Cignoni]{Michele Cignoni$^{1,2}$\thanks{E-mail: michele.cignoni@unibo.it}, Giacomo Beccari$^3$, Angela
    Bragaglia$^2$, and Monica Tosi$^2$
  \\ $^1$ Dipartimento di Astronomia, Universit\`a di Bologna, via Ranzani 1,
     40127 Bologna (Italy)
  \\ $^2$ INAF-Osservatorio Astronomico di Bologna, via Ranzani 1, 40127 Bologna
  (Italy)
  \\ $^3$ European Southern Observatory, Karl Schwarzschild Str. 2, D-85748
  Garching bei M\"unchen (Germany)
  }
\begin{document}

\date{31/03/2011}

\pagerange{\pageref{firstpage}--\pageref{lastpage}} \pubyear{2011}

\maketitle

\label{firstpage}

\begin{abstract}
A comprehensive census of Galactic open cluster properties places
unique constraints on the Galactic disc structure and evolution. In
this framework we investigate the evolutionary status of three
poorly-studied open clusters, Berkeley~31, Berkeley~23 and King~8, all
located in the Galactic anti-centre direction. To this aim, we make
use of deep LBT observations, reaching more than 6 mag below the main
sequence Turn-Off. To determine the cluster parameters, namely age,
metallicity, distance, reddening and binary fraction, we compare the
observational colour-magnitude diagrams (CMDs) with a library of
synthetic CMDs generated with different evolutionary sets (Padova,
FRANEC and FST) and metallicities. We find that Berkeley~31 is
relatively old, with an age between 2.3 and 2.9 Gyr, and rather high
above the Galactic plane, at about 700 pc. Berkeley~23 and King~8 are
younger, with best fitting ages in the range 1.1-1.3 Gyr and 0.8-1.3
Gyr, respectively. The position above the Galactic plane is about
500-600 pc for the former, and 200 pc for the latter. Although a
spectroscopic confirmation is needed, our analysis suggests a
sub-solar metallicity for all three clusters.

\end{abstract}


\section{Introduction}\label{intro}

This paper is part of the BOCCE (Bologna Open Cluster Chemical
Evolution) project, described in detail by \cite{bt06} and aimed at
deriving precise and homogeneous ages, distances, reddenings, and
chemical abundances for a large sample of open clusters (OCs).  Our
final goal is to study the disc of our Galaxy and its formation and
evolution. In fact OCs are among the best tracers of the disc
properties (e.g., \citealt{pt81,friel95}, \citealt*{taat};
\citealt{fbh}).  We have already published results based on photometry 
for 23 OCs
\citep[see][and references therein]{bt06,andreuzzi}, concentrating on
the old ones, the most important to study the early epochs of the
Galactic disc. With the three OCs presented here we have a sample of
20 clusters with ages older than 1 Gyr, i.e. about 10 per cent of all
known old clusters (see the catalogue by \citealt{dias02b} and its web
updates).

The three clusters examined in this paper are Berkeley~23 ($l=192.6^o,
b=5.4^o$), Berkeley~31 ($l=206.2^o, b=5.1^o$), and King~8 ($l=176.4^o,
b=3.1^o$). They are old, distant clusters in the anti-centre
direction. They were selected because they should all lie beyond a
Galactocentric distance of 12 kpc and could then be useful to
understand the nature and properties of the outer Galactic disc. In
particular, they are located in the region where the radial
metallicity distribution seems to change its slope (e.g.,
\citealt{friel2010}; see the discussion in \citealt{andreuzzi}).

These three clusters have already been studied to different degrees in
the past, but with contrasting results. We present here high quality
photometric data, to improve on the previous determinations of their
parameters. As done for all our past work we use comparison of
observational CMDs to synthetic ones generated using different sets of
evolutionary tracks.

Berkeley~23 has been the subject of two photometric
works. \cite{ann2002} observed it with a 1.8m telescope in $UBVI$ as
part of a study of 12 OCs, deriving a reddening $E(B-V)=0.40\pm0.05$
(from the two-colour diagram), a distance modulus
$(m-M)_0=14.2\pm0.3$, a metallicity [Fe/H]=+0.07, and an age of 0.79
Gyr. However, the isochrone they chose does not seem to reproduce the
red clump.  \cite{hasegawa2004} observed 14 OCs with a 65cm telescope,
obtained $BVI$ photometry and derived the clusters' parameters using
isochrones. Their results for Be~23 differ from the previous ones:
$E(B-V)=0.30$, $(m-M)_0=13.81$, metallicity Z=0.004 (equivalent to
[Fe/H]$\simeq-0.7$), and age 1.8 Gyr. \cite{bss}, who presented a
catalogue of Blue Straggler stars in OCs (they have about 1900
candidate BSS in about 430 clusters), find that Be~23 is rich in this
kind of stars.

Berkeley~31 is the best studied of the three OCs. \cite{guetter1993}
published $UBVI$ photometry obtained at the 1m USNO telescope. He
determined the reddening and the metallicity using the two-colour
diagram, and the distance and age with isochrone fit, deriving
$E(B-V)=0.13$, $(m-M)_0=13.6$, [Fe/H]$=-0.4$, and age 8 Gyr.  This old
age is probably an artifact of the outdated stellar
models. Furthermore, the isochrones shown do not reach the red clump,
while the difference in magnitude between it and the main sequence
(MS) turn-off (TO) is a powerful age indicator (for different
definitions of this $\delta V$ see, e.g., \citealt{twa};
\citealt*{pjm94}).  \cite{pjm94} obtained $VI$ photometry, but could
not derive directly $\delta V$; the value they give would put Be~31 at
the same age of M67. \cite{hasegawa2004}, who call it Biurakan~7,
derive from isochrone fit $E(B-V)=0.15$, $(m-M)_0=14.83$, Z=0.008, and
age of 2.2 Gyr.  The discording results are not limited to the
photometry. \cite{friel2002}, using low resolution spectra of 24 stars
(17 defined as members) derived an average radial velocity $<RV>=+41$
(rms=15) km~s$^{-1}$ and a metallicity [Fe/H]=$-0.40$ (rms 0.16)
dex. \cite*{yong} obtained high resolution spectra for five stars;
they notice that the RV dispersion is very high for an OC, maybe
because they observed by chance a large fraction of binaries or
because their sample includes field stars with RV similar to the one
of Be~31. They were able to perform an abundance analysis only on one
star and derived [Fe/H]$=-0.57$. A similar situation was found by
\cite{friel2010} who obtained high resolution spectra of another two
stars, with discrepant RVs -differing also from the values by
\cite{yong}- and metallicity (although the two values, [Fe/H]=$-0.22$
and $=-0.32$ are not too distant). They could not determine which of
the two stars, if any, is a true cluster member. They also noticed
that the published proper motions \citep*{dias02a} should be of field
stars in the direction of the cluster, given their bright magnitudes.
Finally, also for Be~31 \cite{bss} found a large number of BSS.

The first photometry (in the $GRU$ filters) of King~8 was published by
\cite{svol}, who described a young cluster, with an age of about 5
Myr, $E(B-V)=0.34$, and $(m-M)_0=13.3$. \cite{christian1981} presented
BV photographic photometry and low resolution spectra. She gave
$E(B-V)\simeq0.7$ (with a possible differential reddening up to 0.1
mag), Z$<0.001$, distance from the Sun of 3-4 kpc, and age 0.8
Gyr. These values were partially changed by \cite{christian1984},
where $E(B-V)=0.55$ and [Fe/H]$\simeq-0.4$ or $-0.5$ dex. These two
works were later used by \cite{taat} to derive a larger distance,
$(m-M)=15.30$, adopting the metallicity [Fe/H]$=-0.46$ of
\cite{fj83}, based on two stars observed at low
resolution. \cite*{koposov} re-discovered King~8 in their automated
search for new OCs on the 2MASS images. They derived the cluster
parameters using isochrones in the $J,J-H$ plane, finding
$E(B-V)=0.44$, distance from the Sun of 2.4 kpc, and age 1.1 Gyr. They
also noted that their numbers differ from what is listed in the
\cite{dias02b} catalogue.

In summary, even if apparently these three OCs have already been
studied, their properties are still insecure. Be~23 could be rather
young and metal rich (\citealt{ann2002}: 0.79 Gyr, [Fe/H]=+0.07) or
rather old and metal-poor (\citealt{hasegawa2004}: 1.8 Gyr,
[Fe/H]=$-0.7$).  Be~31 could be very old or of intermediate age, and
at very different Galactocentric radii (\citealt{guetter1993}: 8 Gyr,
13 kpc; \citealt{hasegawa2004}: 2.2 Gyr, 17 kpc), while there is some
concordance on $E(B-V)$ around 0.13-0.15 and metallicity around $-0.4$
dex (but with caveats on membership). King~8 has probably an age
around 1 Gyr and sub-solar metallicity but the reddening and distance
are uncertain (\citealt{taat}: $E(B-V)$=0.55, d$_\odot$=5.24 kpc;
\citealt{koposov}: $E(B-V)$=0.44, d$_\odot$=2.4 kpc).

In the present paper we will describe our new observations and the
resulting CMDs (Sect. 2), the radial extension of the clusters
(Sec. 3), the derivation of their age, distance, reddening, and
metallicity using comparison to synthetic CMDs (Sect. 4). A discussion
and summary is presented in Sec. 5.

\section{The data}\label{data}

\subsection[]{Observations}
\label{obs}
The three clusters were observed in service mode at the Large
Binocular Telescope (LBT) on Mt. Graham (Arizona) with the Large
Binocular Camera (LBC) in 2008.  There are two LBCs, one optimised for
the UV-blue filters and one for the red-IR ones, mounted at each prime
focus of the LBT.  Each LBC uses four EEV chips (2048$\times$4608
pixels) placed three in a row, and the fourth rotated $90\deg$ with
respect to the others. The field of view of LBC is equivalent to
23\arcmin$\times$23\arcmin, with a pixel sampling of 0.23\arcsec.  For
technical reasons, only LBC-Blue was available at the time of our
observations, so we collected only $B$ and $V$ data.  The clusters
were positioned in the central chip (\# 2) of the LBC-Blue CCD
mosaic. Table~\ref{oss} gives a log of the observations.
\begin{table*}
\centering
\caption{Log of observations for the LBT and TNG, respectively. The
  exposure times are in seconds.}
  \setlength{\tabcolsep}{5.0mm}
\begin{tabular}{lccccc}
\hline\hline
Cluster & RA & Dec &Date & B & V  \\
& (2000) &(2000) & &exptime & exptime \\
\hline
\multicolumn{6}{c}{LBC@LBT}\\
Be~23 &06 33 15 &+20 34 50 & 02 Dec 2008 &1s, 3x5s, 4x90s&1s, 3x5s, 3x60s\\
Be~31 &06 57 37 &+08 21 18 & 02 Dec 2008 &1s, 3x5s, 3x90s&1s, 3x5s, 3x60s\\
King~8 &05 49 24 &+33 40 60 & 02 Dec 2008 &1s, 3x5s, 4x90s&1s, 3x5s, 3x60s\\
\multicolumn{6}{c}{DOLORES@TNG}\\
Be~23 &06 33 15 &+20 31 30 & 03 Jan 2009 &10s, 20s & 5s, 10s, 20s  \\
Be~31 &06 57 37 &+08 17 20 & 03 Jan 2009 &10s, 20s &10s, 20s \\
King~8 &05 49 18 &+33 37 50 & 03 Jan 2009 &10s, 20s &10s, 20s  \\
\hline
\end{tabular}
\label{oss}
\end{table*}
While the seeing was good (below 1\arcsec \ for all images, and below
0.8\arcsec \ for many), the observing conditions
were not photometric. To calibrate our photometry we obtained a few
shallow images using DOLORES (Device Optimised for for the LOw
RESolution) at the Italian Telescopio Nazionale Galileo (TNG) on a
night dedicated to another programme and to service observations. For
this instrument the field of view is $8.6\arcmin\times8.6\arcmin$,
with a scale of 0.252 \arcsec/pix. The seeing was bad (about 3\arcsec) but this is not a problem since the fields are
not crowded. Only the central part of the clusters was observed,
barely reaching stars at the turn-off of the main sequence. The two
standard star fields PG0918+029 and PG1323-085 from
\cite{landolt} were also obtained.

\begin{figure*}
\begin{center}
\includegraphics[width=16cm]{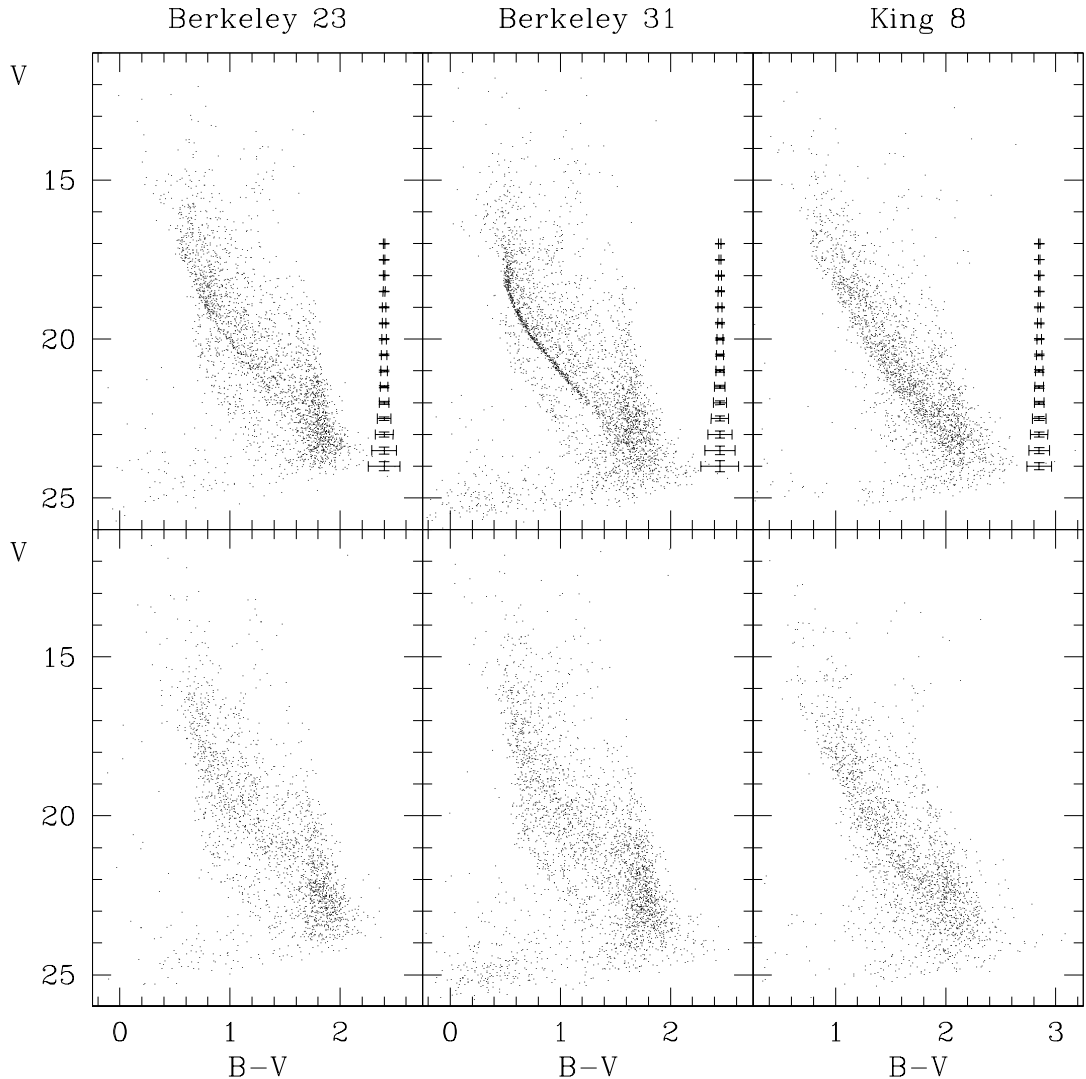} 
\caption{Upper panels: CMDs of Be~23, Be~31, and King~8 (chip \#2).
  Lower panels: CMDs of the relative comparison fields (chip
  \#1). Photometric errors, as derived from the artificial star tests,
  are shown on the right side of the CMDs.}
\label{cmdall}
\end{center}
\end{figure*}

\subsection[]{Data reduction}
\label{redu}
Given the large area covered by LBC and the small angular dimensions
of our targets (see Sections \ref{prof}, \ref{fit}) we only reduced
two of the four chips, the one centered on the cluster (chip \#2) and
a second one to be used as a comparison to separate the cluster from
the field stars (chip \#1).

The raw LBC images were corrected for bias and flat field, and the
overscan region was trimmed using a pipeline specifically developed
for LBC image prereduction by the Large Survey Center team at the
Rome Astronomical Observatory\footnote{http://lsc.oa-roma.inaf.it/}.
The source detection and relative photometry was performed
independently on each B and V image, using the PSF-fitting code
{ DAOPHOTII/ALLSTAR} \citep{stetson87,stetson94}. The brightest stars,
saturated in the deepest images, where efficiently recovered through
the accurate magnitude measures from the short exposure time
images. The average and the standard error of the mean of the
independent measures obtained from the different images were adopted
as the final values of the instrumental magnitude and 
uncertainty.

More than 300 stars from the {\it Guide Star Catalogue 2.3} where
used as astrometric standards to find an accurate astrometric solution
to transform the instrumental positions, in pixels, into J2000
celestial coordinates. To this aim we adopted the code CataXcorr,
developed by Paolo Montegriffo at INAF - Osservatorio Astronomico di
Bologna, and successfully used by our group in the past 10 years. The
r.m.s. scatter of the solution was $\sim0.3\arcsec$ in both RA and
Dec.

We derived the completeness level of the photometry by means of
extensive artificial stars experiments following the recipe described
in \cite{bel02}.  More than 150000 artificial stars have been
uniformly distributed on the chip sampling the cluster in groups of
9000 stars at a time, to avoid changing the crowding conditions. The
whole data reduction process has been repeated as in the real case and
the fraction of recovered stars was estimated at each magnitude
level. We also checked that the completeness does not change from the
central to the external parts.  Results for the three clusters are
presented in Table~\ref{compl}.

With the standard stars observed at the TNG we derived the following
calibration equations
$$ B=b+0.1024\times(b-v)+1.3580 ~~~(rms=0.0118)$$
$$ V=v-0.0506\times(b-v)+1.0416 ~~~(rms=0.0167)$$ where $B,V$ are the
calibrated magnitudes and $b,v$ the instrumental ones (after aperture
correction). Finally, the deep LBT catalogues were cross-correlated
with the calibrated, shallow TNG ones and the stars in common were
used as ``secondary standards" to extend the calibration to the whole
catalogues, both for the central and the external chip.

We produced three catalogues with identification, equatorial
coordinates, $B$ and $V$ magnitudes, with errors containing about
3600, 6500, and 4000 stars for Be~23, Be~31, and King~8,
respectively. These catalogues will be made public through the
WEBDA\footnote{http://www.univie.ac.at/webda/webda.html}.

Fig.~\ref{cmdall} shows the resulting CMDs from our photometry for the
central and comparison field chips (upper and lower panels,
respectively). In the upper panels we also indicate the photometric
errors in magnitude and colour as derived from extensive artificial
star tests. They range from less than 0.01 mag at the bright limit to
less than 0.1 mag around $V=24$. The three OCs are visible over the
field background, but they represent a minority of the stars in each
frame. In the most favourable case, Be~31, the main sequence is rather
tight and clearly extended down to $V \approx 23-24$, while for Be~23
and King~8 the MS is recognisable from the field with some
difficulty. A discussion of the field population sampled in these CMDs
is presented in Sect. 5.

\begin{table*}
\centering
\caption{Completeness of the photometry for the three clusters.}
\setlength{\tabcolsep}{5.0mm}
\begin{tabular}{c cc cc cc}
\hline\hline
mag &compl B & compl V & compl B & compl V & compl B & compl V\\
&\multicolumn{2}{c}{Be 23} &\multicolumn{2}{c}{Be 31} &\multicolumn{2}{c}{King 8} \\
\hline
17.25  & 1.00 $\pm0.11$ & 0.99 $\pm0.05$ & 0.95 $\pm0.11$ & 0.96  $\pm0.06$ &  0.98 $\pm0.05$  & 0.99 $\pm0.02$ \\
17.75  & 1.00 $\pm0.11$ & 0.98 $\pm0.04$ & 0.95 $\pm0.12$ & 0.97  $\pm0.06$ &  0.98 $\pm0.05$  & 0.99 $\pm0.04$ \\
18.25  & 0.99 $\pm0.06$ & 0.98 $\pm0.04$ & 0.98 $\pm0.09$ & 0.98  $\pm0.05$ &  0.99 $\pm0.04$  & 0.99 $\pm0.03$ \\
18.75  & 0.98 $\pm0.04$ & 0.97 $\pm0.03$ & 0.97 $\pm0.06$ & 0.97  $\pm0.05$ &  0.99 $\pm0.04$  & 0.98 $\pm0.03$ \\
19.25  & 0.97 $\pm0.04$ & 0.97 $\pm0.03$ & 0.98 $\pm0.05$ & 0.95  $\pm0.04$ &  0.98 $\pm0.03$  & 0.97 $\pm0.03$ \\
19.75  & 0.98 $\pm0.03$ & 0.97 $\pm0.03$ & 0.97 $\pm0.05$ & 0.96  $\pm0.04$ &  0.98 $\pm0.03$  & 0.97 $\pm0.03$ \\
20.25  & 0.97 $\pm0.03$ & 0.96 $\pm0.03$ & 0.96 $\pm0.04$ & 0.95  $\pm0.04$ &  0.98 $\pm0.03$  & 0.97 $\pm0.03$ \\
20.75  & 0.97 $\pm0.03$ & 0.95 $\pm0.03$ & 0.96 $\pm0.04$ & 0.94  $\pm0.04$ &  0.98 $\pm0.03$  & 0.96 $\pm0.02$ \\    
21.25  & 0.98 $\pm0.03$ & 0.94 $\pm0.03$ & 0.95 $\pm0.04$ & 0.92  $\pm0.04$ &  0.97 $\pm0.03$  & 0.95 $\pm0.02$ \\    
21.75  & 0.96 $\pm0.03$ & 0.93 $\pm0.03$ & 0.94 $\pm0.04$ & 0.90  $\pm0.03$ &  0.96 $\pm0.03$  & 0.95 $\pm0.02$ \\    
22.25  & 0.95 $\pm0.03$ & 0.90 $\pm0.02$ & 0.94 $\pm0.04$ & 0.89  $\pm0.03$ &  0.96 $\pm0.02$  & 0.92 $\pm0.02$ \\    
22.75  & 0.94 $\pm0.03$ & 0.87 $\pm0.02$ & 0.90 $\pm0.03$ & 0.84  $\pm0.03$ &  0.95 $\pm0.02$  & 0.90 $\pm0.02$ \\    
23.25  & 0.91 $\pm0.02$ & 0.78 $\pm0.02$ & 0.88 $\pm0.03$ & 0.80  $\pm0.03$ &  0.93 $\pm0.02$  & 0.88 $\pm0.02$ \\    
23.75  & 0.89 $\pm0.02$ & 0.55 $\pm0.02$ & 0.85 $\pm0.03$ & 0.72  $\pm0.02$ &  0.93 $\pm0.02$  & 0.80 $\pm0.02$ \\    
24.25  & 0.85 $\pm0.02$ & 0.29 $\pm0.01$ & 0.82 $\pm0.03$ & 0.53  $\pm0.02$ &  0.89 $\pm0.02$  & 0.59 $\pm0.01$ \\    
24.75  & 0.74 $\pm0.02$ & 0.10 $\pm0.01$ & 0.73 $\pm0.02$ & 0.26  $\pm0.01$ &  0.83 $\pm0.02$  & 0.26 $\pm0.01$ \\    
\hline				    	      
\end{tabular}
\label{compl}
\end{table*}

\begin{figure}
\begin{center}
\includegraphics[bb=50 200 580 500, clip,width=8.6cm]{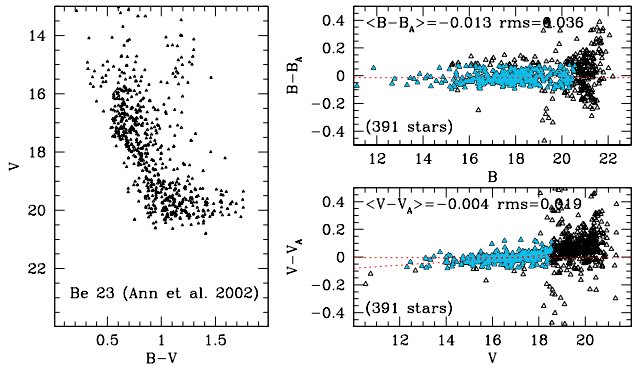} 
\caption{CMD of Be~23 by Ann et al. (2002) on the same scale used for
  next figures with our data. Right panels: differences between our
  magnitudes and theirs, in $B$ (upper panel) and $V$ (lower
  panel). Open symbols indicate all stars in common, light blue filled
  ones the stars within 2$\sigma$ from the average, used to compute
  the mean differences. }
\label{confbe23}
\end{center}
\end{figure}

\begin{figure}
\begin{center}
\includegraphics[bb=50 200 580 500, clip,width=8.6cm]{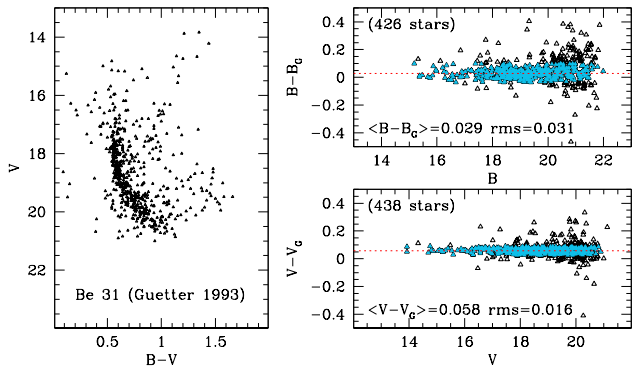} 
\caption{Left panel: CMD of Be~31 by Guetter (1993) and comparison
  with our data (see previous figure)}
\label{confbe31}
\end{center}
\end{figure}

\begin{figure}
\begin{center}
\includegraphics[bb=50 200 580 500, clip,width=8.6cm]{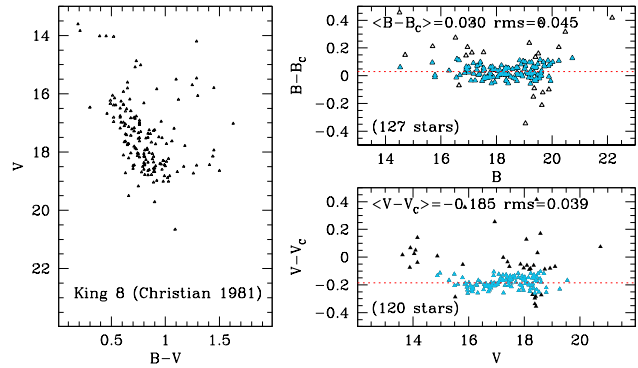} 
\caption{Left panel: CMD of King~8 by Christian (1981) and comparison 
 with our data (see previous figures). }
\label{confking8}
\end{center}
\end{figure}

\begin{figure}
\centering
\includegraphics[width=7.5cm]{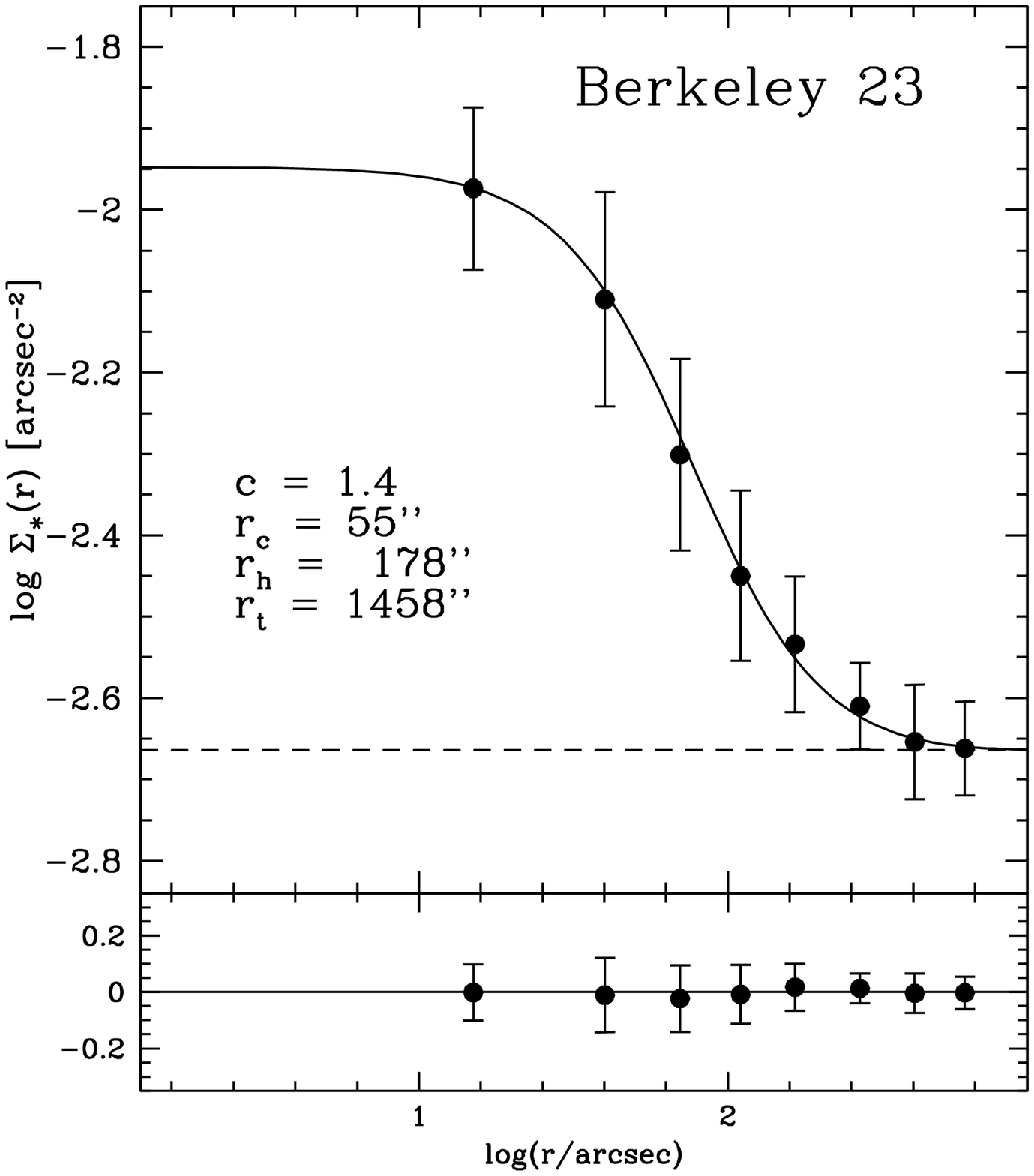}
\includegraphics[width=7.5cm]{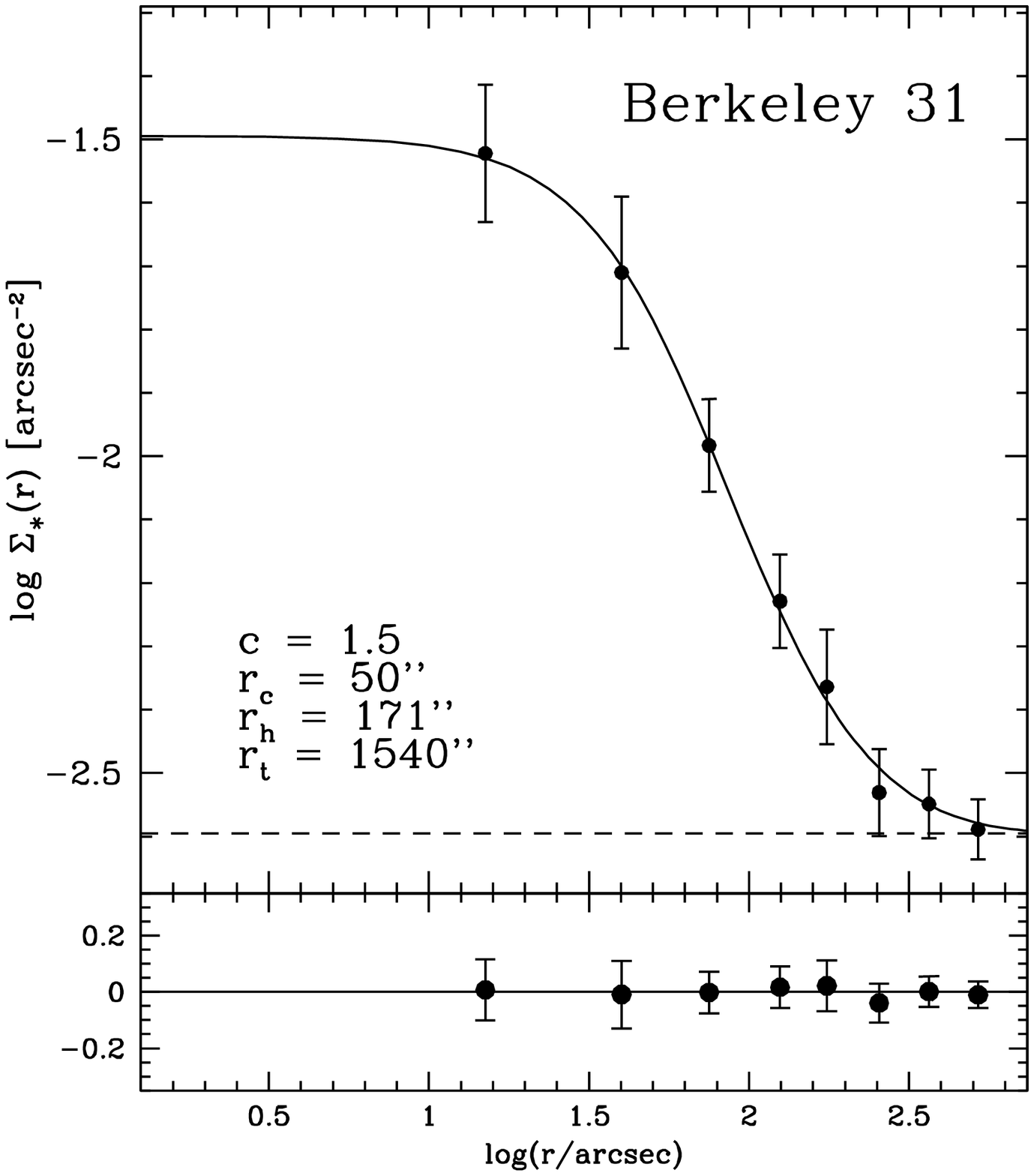}
\includegraphics[width=7.5cm]{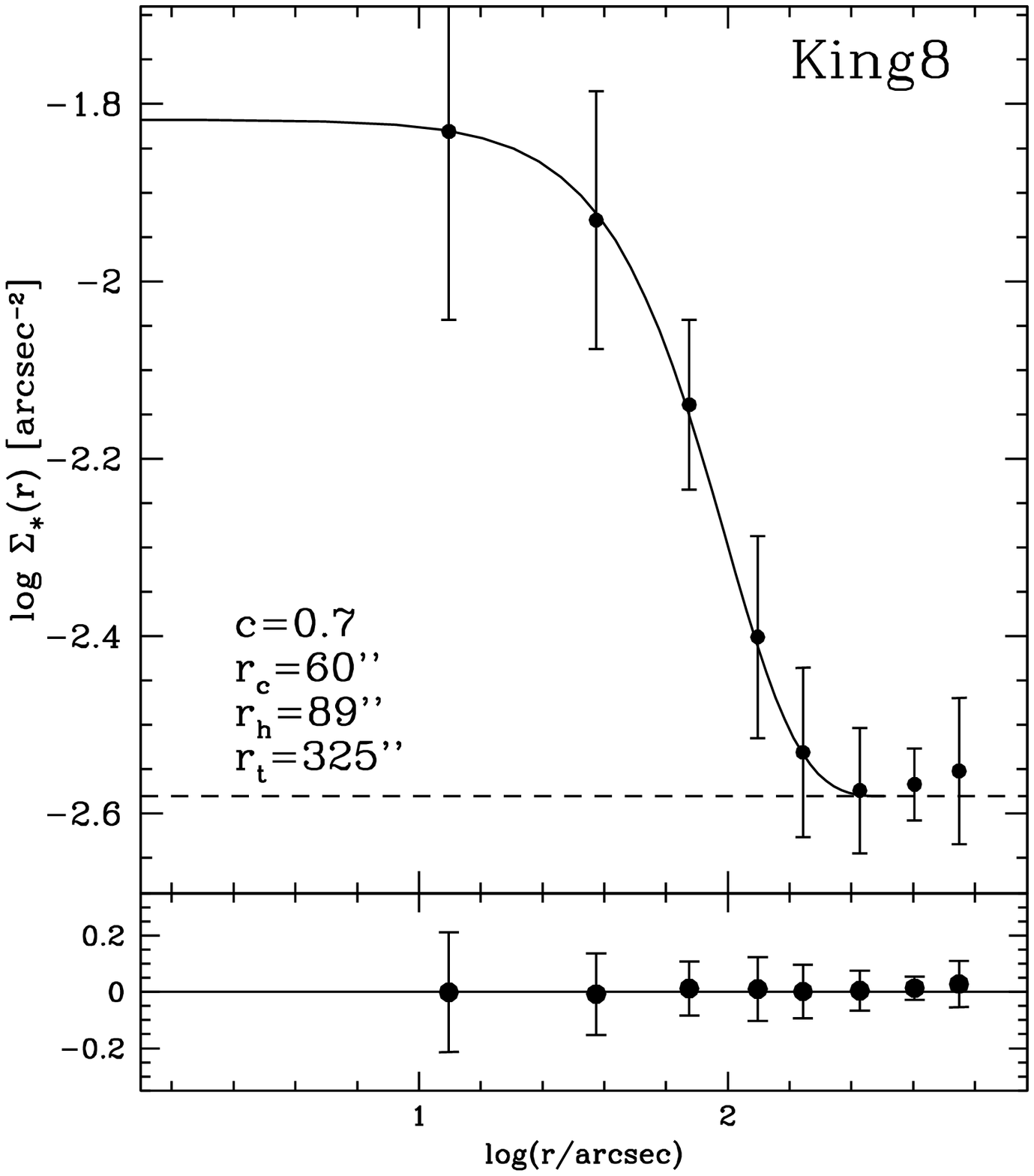}
\caption{Observed surface density profiles (filled circles and error
  bars) in units of number of stars per square arc-second. The solid
  lines are the King models that best fit the observed density profile
  over the entire cluster extension.  }
\label{profile}
\end{figure}

\subsection[]{Comparison to previous data}\label{comp}

We compare our photometry for the three OCs with \cite{ann2002} for
Be~23, \cite{guetter1993} for Be~31, and \cite{christian1981} for
King~8. We do not consider the photometry of Be~31 by \cite{pjm94}
because it has only $V$ and $I$ and that of \cite{hasegawa2004} of
Be~31 and Be~23 because they made public through the WEBDA only the
bright stars ($V\le18$).

We downloaded the literature photometry files from the WEBDA site and
cross-correlated our catalogues with them. Figs~\ref{confbe23},
\ref{confbe31}, and \ref{confking8} show the results of this
comparison between our and literature magnitudes.

For Be~23 (Fig.~\ref{confbe23}) the difference in the $B$ filter are
very small (on average -0.013 mag), while the $V$ shows a small trend
with magnitude. We do not know if this is due to our photometry or
theirs; however, no trend is present for the two other OCs, that were
calibrated with the same equations, and this is in favour of the
latter possibility. In the same figure we also show the CMD obtained
by \cite{ann2002} to give an immediate impression of the much better
quality of our data (see Fig.~\ref{cmdall}).

In the case of Be~31 (Fig.~\ref{confbe31}), we find small average
differences both in $B$ and $V$ (0.029 and 0.058 mag, respectively),
and no trends. The CMD by \cite{guetter1993}, even if of good quality,
is poorer than ours, which reaches fainter and shows details
(e.g. near the MS TO) not visible in Guetter's.

Finally, King~8 (Fig.~\ref{confking8}) has only the photographic data
by \cite{christian1981}. While the difference between the two $B$
values are small (-0.030 mag), for $V$ we reach an average -0.185
mag. Again, this cluster was calibrated exactly as the other two and,
given the much smaller differences with completely independent sources
found for the other OCs, we think that the problem is (mostly) in the
old photographic photometry.

\section{The radial profiles}\label{prof}

Taking advantage of the combination of the wide field and deep imaging
capabilities of LBC on LBT, we determined the cluster structural
parameters from the star density profiles.  As a first step, the
centre of gravity ($C_{grav}$) was estimated simply by averaging the
$\alpha$ and $\delta$ coordinates of clusters stars. In order to avoid
completeness effects and strong field contamination, we considered
samples with different limiting magnitudes ($V<20,20.5,21$) and with
colour $B-V<1.5$. For each sample we iteratively calculated the
barycentre of the stars~\citep[see][]{be06} inside a circular area of
70\arcsec radius.  The size of the area, somehow arbitrary, allows us
to sample a gradient in the stellar radial distribution (see details
later on this paragraph) and to minimise field contamination when not
removed after colour and magnitude cuts.  The centre of gravity has
been calculated as the average of the measures in the three magnitude
ranges and the standard deviation as the related error. The values of
$C_{grav}$ and relative errors are reported on Table~\ref{tab_param},
where we also indicate the coordinates found in the WEBDA, since they
are slightly different.

The projected density profile was determined for each cluster using
direct star counts following the procedure described
in~\cite{la10}. Shortly, the photometric catalogue is divided in
concentric annuli centered on the cluster $C_{grav}$.  Each annulus is
then divided in at least three angular sectors. The density of the
annulus is calculated as the mean of the stellar density in each
sector with the standard deviation of the mean used as the associated
error. The observed radial density profile for each target cluster is
shown in Fig. \ref{profile}, with the abscissas corresponding to the
mid-point of each radial bin. Notice that, in order to minimise field
contamination, we used only stars with magnitudes $V<21$ and colour
$B-V<1.5$.

The cluster structural parameters can be estimated through a best
fitting procedure of the derived density profile with a
suitable~\cite{k66} model~\citep[as done, e.g., in][]{so10}.  An isotropic,
single-mass King model projected onto the cluster area was computed
adopting the~\cite{sp95} code.  The model producing the best fit for
each cluster is shown in Fig. \ref{profile}, together with the
structural parameters core radius ($r_c$), half-mass radius ($r_h$),
tidal radius ($r_t$) and concentration ($c=\ln(r_t/r_c)$, all indicated in
Table~\ref{tab_param}.  The residual of the fit of the model with
each observed point is shown in the lower panel of each density plot.
Notice that the estimate of the background level is based on the star
counts in the peripheral regions where the number of field stars is
significantly higher than that of cluster stars.

\begin{table*}
\centering
\caption{New clusters structural parameters; numbers in parenthesis are the errors on RA and Dec.}
\begin{tabular}{lcccccccc}
\hline\hline
Cluster & RA & Dec & RA & Dec & $r_c$ & $r_h$ & $r_t$ & $c$ \\
        &\multicolumn{2}{c}{present paper} &\multicolumn{2}{c}{WEBDA} &&&&\\
\hline
Be~23  &06 33 15 (1.24\arcsec) &20 31 57 (1.98\arcsec) &06 33 30 &20 33 00 & 55\arcsec & 178\arcsec & 1458\arcsec & 1.4 \\
Be~31  &06 57 37 (2.84\arcsec) &08 18 20 (2.91\arcsec) &06 57 36 &08 16 00 & 50\arcsec & 171\arcsec & 1540\arcsec & 1.5\\
King~8 &05 49 18 (3.21\arcsec) &33 37 50 (2.22\arcsec) &05 49 24 &33 38 00 & 60\arcsec & 89\arcsec  & 325\arcsec & 0.7 \\
\hline
\end{tabular}
\label{tab_param}
\end{table*}

\section{Clusters' parameters via synthetic colour-magnitude diagrams}
\label{fit}

In order to determine age, metallicity, distance, mean Galactic
reddening, and binary fraction we have compared the observational CMD with
a library of artificial populations Monte Carlo generated (see
\citealt{Tosi2007}). Different sets of stellar tracks\footnote{The
  same used in our past works, to maintain homogeneity: the old Padova
  \citep{bbc93,bbc94}, the FRANEC \citep{franec}, and the FST ones
  \citep{fst}. See \cite{bt06} for a description of their main
  properties and a detailed justification of their use even if newer
  tracks have appeared in the meantime.} have been applied, fitting
both primary age-sensitive descriptors, like the luminosity of the MS
reddest point (``red hook'', RH), the red clump (RC) and the MS
termination point (MSTP, evaluated at the maximum luminosity reached
after the overall contraction, OvC, and before the runaway to the
red), and secondary CMD features (sensitive to a broad range of
parameters) like the RH colour, the luminosity at the base of the red
giant branch (RGB), the RGB colour and inclination, the RC colour. The
most valuable age indicator is in principle the TO point, evaluated at
the bluest point after the overall contraction; however, for young and
intermediate age clusters this phase is poorly populated. More in
general, colour constraints are less reliable than luminosity
constraints, since the former are much more influenced by theoretical
uncertainties like colour transformations and the super-adiabatic
convection.

In order to make a meaningful comparison, all synthetic CMDs are
combined with stars picked from an equal area of the adjacent field
which is located about $9^{\prime}$ away from the cluster centre. For
each cluster, the synthetic CMD is populated until the total number of
stars (synthetic plus field) equals the observed number of cluster
stars brighter than $V=20$ and corrected for photometric errors and
incompleteness, as derived from the artificial star tests.

As a first step, we seek cluster parameters that are common to all
solutions. Differential reddening and fraction of binaries are the
typical cases, because they are almost independent of the specific set
of stellar tracks. When these quantities are fixed (essentially
matching the MS width), the magnitude difference between RH, MSTP and
RC is effectively used to constrain the age (see e.g., Fig. 8 in
\citealt{castellani2003}). Once the best fitting age is established,
all remaining parameters (mean Galactic reddening, metallicity, and
distance) are varied till the colour of the MS and the magnitudes of
RH, MSTP, RC are matched. A final test of the fit quality considers
the RH, RGB and RC colours as well as the luminosity at the base of
the RGB.

Critical points of this process are the MS and the RC fitting. The
former issue involves the upper and the lower ends of the MS. The RH
morphology can vary from a vertical orientation to a very hooked
morphology with the change of age, metallicity, overshooting,
microscopic physics, etc., while the inclination of the lower MS
follows the equation of state and the adopted colour
transformations. In this context, a combination of parameters is
defined ``acceptable'' when the corresponding synthetic CMD fits
``most'' of the visible MS shape. On the other hand, the RC fitting
can be hampered by both theoretical and statistical uncertainties. The
RC morphology and luminosity strongly depend on fundamental parameters
like the metallicity, the age and the helium content, as well as by
more subtle physical inputs like the efficiency of core overshooting
and the amount of mass loss in the pre He burning phase (see
\citealt{castellani00} for an in-depth discussion). Moreover the
number of RC stars is rather small in all three clusters, increasing
the probability of confusion with RGB and field stars and not allowing
very precise constraints on the derived age.

In the following, we present the three clusters from the oldest to the
youngest one.  Note that we did not make use of the existing
information on RVs for Be~31 because they are inconclusive regarding
the membership (see Sect. 1 and the discussion in \citealt{friel2010})
and for Be~23 because they are limited to two stars.

\subsection[]{Berkeley 31}

\begin{figure*}
\begin{center}
    \includegraphics[width=14cm]{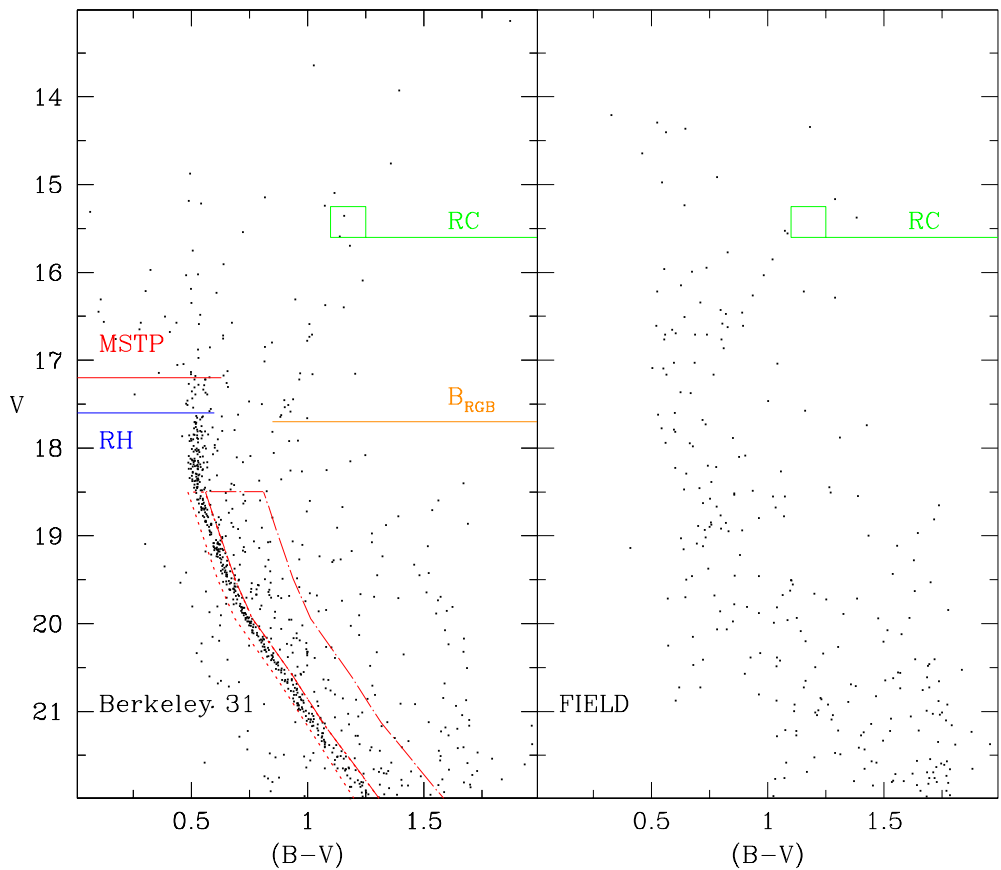} 
    \end{center}
  \caption{Left panel: CMD of stars within $2.5^{\prime}$ of the
    centre of Be~31. We indicate the luminosity level of the RH, the
    MSTP, the RC and the base of the RGB (BRGB). The dotted and the
    dot-dashed boxes are, respectively, used to estimate the fraction
    of single and binary stars. Right panel: CMD of field stars (as
    evaluated from an equal area adjacent region).}
  \label{be31_2p5}
\end{figure*}

In this group of OCs, Be~31 is the richest in term of members. The
contamination of field stars is evident, especially above the
sub-giant branch and blue-ward of the RGB. In order to reduce the
number of field interlopers, but still retain a significant number of
cluster members, we restrict our sample to the stars within a radius
of 2.5\arcmin \ from the cluster centre (that is 2.5 times $r_c$,
slightly more than $r_h$, and well within $r_t$, see
Fig.~\ref{profile} and Table~\ref{tab_param}).  The corresponding CMD
is used for the determination of the cluster parameters. The most
important features are emphasised in the left panel of Fig. \ref{be31_2p5} (while the right panel shows the control field CMD).
 First of all, the OvC gap (between $V=17.45$ and $V=17.6$) and the
 base of the RGB (around $V=17.65$) are clearly visible. The region of
 the RC is not well populated, but it is at least identifiable with
 the mild, but in our view significant, excess\footnote{Such an excess
   is found with respect to other equal-area CMD randomly chosen from
   the control field.} of stars in the magnitude range $15<V<15.6$
 and colour $B-V\approx 1.1-1.2$.

Finally, an excess of stars running parallel and red-ward to the MS
indicates that a fraction of stars is in binary systems.
 
A remarkable number of stars is located blue-ward ($B-V<$0.4) and
mainly above ($15<V<17.5$) the MSTP (see Fig. \ref{be31_2p5}), which
cannot be explained by field stars (the control field does not contain
any such blue object). As already suggested by \cite{bss} these
objects are likely blue stragglers.

As already found several times in the past, the MS is broader than we
would have expected from the photometric errors; what is surprising is
that such an effect is larger around the RH. This might be a result of
the combined effect of differential reddening and RH morphology.  On a
more speculative side, keeping in mind the strong dependence on age of
the RH morphology, it is also conceivable that the RH colour width
might represent a signature of a prolonged star formation.

To solve the problem of the MS scatter, one has to conclude that
either a differential reddening of about $E(B-V)=0.05$ is present or a
large fraction of binaries hosts very low mass companions (the
photometric properties of a binary system are very close to the one of
the primary alone, if the primary is below $2\,M_{\odot}$ and the
companion's mass is less than half the primary's mass; see e.g. Fig. 1
in \citealt{Hurley1998}), or both. Given the patchy nature of the
obscuring material in the Galactic thin disc we are more inclined to
the reddening hypothesis. In addition to the mean Galactic reddening,
in the following we will adopt a differential reddening\footnote{Here
  the total reddening is considered a sum of a mean Galactic component
  and a random component (``differential'') which includes the
  foreground fluctuation and a possible internal reddening.} of about
0.05 mag unless otherwise stated.

A rough estimate of the binary fraction can be obtained by evaluating
the number of stars in two appropriate CMD boxes, one along the MS (see
the dotted line in Fig. \ref{be31_2p5}) and one red-ward of the MS
(chosen to cover the binary sequence; see the dot-dashed line). We
apply magnitude cuts and keep only stars below the RH and brighter
than $V=22$ to avoid evolved stars where the binary evolution can be
quite complicated. To remove the field contamination we have
statistically subtracted the contribution of field stars using several
equal area regions randomly chosen from the control field. The net
result is a binary fraction between 22\% and 26\%. Strictly speaking
these estimates may be slightly underestimated with respect to the
real fraction, since low mass ratio binaries, whose properties are
close to those of single stars, might be partially missed. However, a
mean fraction of 24\% appears to be a reasonable ``ansatz'' and will
be assumed for all the simulations.

Keeping the binary fraction and the differential reddening fixed, we
have searched the best combination of parameters for each set of
tracks and metallicity. Table \ref{t1} summarises the results and the
residual discrepancies. Synthetic RGB, RC, and lower MS colours together
with the magnitude of the RGB base are labelled as ``TB'', ``TR'',
``TF'', or  ``OK'' when they are too blue, too red, too
faint,  or  in agreement with respect to the data.
\begin{table*}
 \centering
  \caption{Be~31 best fit parameters (age, mean Galactic reddening,
    differential reddening, and distance modulus) for various sets of
    models and metallicities. Columns from 7 to 10 indicate how the
    models reproduce the following CMD features: RGB colour, magnitude
    at the base of the RGB, RC colour, and lower MS
    colour. Additionally, the 10th column reports the magnitude below
    which the synthetic MS diverges from the data MS. See text for
    definition of the flags TR, TB, TF, OK.}
  \setlength{\tabcolsep}{1.5mm}
  \begin{tabular}{@{}llllllllll@{}}
  \hline
   Set&Z&Age (Gyr) & $E(B-V)_{\mathrm{M}}$ & $E(B-V)_{\mathrm{D}}$ &$(m-M)_0$&RGB (C) & RGB base (M)& RC (C)&Low MS (C)\\
 \hline
 FST ($\eta=0.2$)&0.02  &2.7   &0.020 &$\pm 0.00$    &14.90 &TR &OK &TB &TB(V$>$19)\\
 FST ($\eta=0.2$)&0.01  &2.8   &0.095 &$\pm 0.025$   &14.60 &OK &OK &TB &TB(V$>$19)\\
 FST ($\eta=0.2$)&0.006 &2.6   &0.165 &$\pm 0.025$   &14.60 &OK &OK &TB &TB(V$>$19)\\

 Padova&0.02  &2.8   &0.00 &$\pm 0.00$   &14.75 &TR &TF &OK &TB(V$>$20.5)\\
 Padova&0.008 &2.6   &0.135 &$\pm 0.025$   &14.45 &TR &OK &TB &TB(V$>$20.5)\\
 Padova&0.004 &2.9   &0.185 &$\pm 0.025$   &14.40 &OK &OK &TB &TB(V$>$20.5)\\

 FRANEC&0.02  &2.5   &0.025&$\pm 0.025  $ &14.80 &TB &TF &TR &OK(V$>$21)\\
 FRANEC&0.01  &2.3   &0.145&$\pm 0.025  $ &14.60 &TR &OK &OK &OK(V$>$21)\\
 FRANEC&0.006 &2.5   &0.175&$\pm 0.025  $ &14.46 &OK &OK &TB &OK(V$>$21)\\
\hline
\end{tabular}
\label{t1}
\end{table*}

Concerning the age, the overall interval of confidence is between 2.3
and 2.9 Gyr. As expected, to reproduce the observed colours, models of
progressively lower metallicity require a larger mean reddening and a
larger distance modulus, because they are intrinsically bluer and
brighter. However, the quality of the fit subtly depends on the
adopted metallicity.

The FST models with $Z=0.006$ and $Z=0.01$ fit reasonable well the RH,
MSTP and RC luminosity levels, the colour of the RGB and the magnitude
of the RGB base but they predict a ``too hooked'' RH (that doesn't
mean the colour is wrong, but rather that the MS shape is too curved
before the RH point), a synthetic MS (below the 19th magnitude) and a
RC bluer than actually observed. In term of cluster parameters, the
former metallicity would suggest a cluster age of 2.6 Gyr,
$E(B-V)=0.165\pm 0.025$ and a distance modulus $(m-M)_{o}=14.6$, the
latter a cluster age of 2.8 Gyr, $E(B-V)=0.095\pm 0.025$ and a
distance modulus $(m-M)_{o}=14.6$. Raising the metallicity to $Z=0.02$
worsens significantly the quality of the fit. In fact, the magnitude
difference between RC and MSTP requires an age of 2.7 Gyr and any
attempt to reconcile the predicted and observed RH colour and
magnitude runs into two additional problems: the reddening should be
lowered to an untenable $E(B-V)\approx 0.02$ (compared to the
\citealt{sfd98} estimate of 0.145) and the synthetic RGB would be too
red.

With the Padova stellar tracks, we find that the lower metallicity,
$Z=0.004$, provides the best match, although suffering from the same
pathologies noticed with the $Z=0.006$ FST set, i.e., too blue
synthetic RC and lowest MS. In this case the estimated age raises to
2.9 Gyr, and we find a mean reddening $E(B-V)=0.185 \pm 0.025$ and a
distance modulus $(m-M)_{o}=14.4$. The metallicity $Z=0.008$ does not
improve the fit and makes the match worse, producing a too red
RGB. Accepting these discrepancies we get an estimated age of 2.6 Gyr,
$E(B-V)=0.135 \pm 0.025$ and distance modulus
$(m-M)_{o}=14.45$. Finally, the Padova $Z=0.02$ set is not acceptable
for the same reason as the solar FST: matching the MSTP, RC and the RH
magnitudes would require a null reddening, contrary to \cite{sfd98}
values.

The FRANEC models are the only ones able to reproduce well the faint
end of the MS (down to $V\approx\,20-20.5$). Using the metallicity
$Z=0.006$ we can reproduce all important features apart for a mismatch
in the colour of the RC that is too blue. In this case the best
combination of parameters is: cluster age 2.5 Gyr, $E(B-V)=0.175\pm
0.025$, and distance modulus $(m-M)_{o}=14.46$. The metallicity
$Z=0.01$ reduces the RC colour discrepancy, but to the cost of a
synthetic RGB that is too red by as much as $\sim\,0.1$ mag and
under-luminous at the base by as much as $\sim\,0.25$. Using $Z=0.01$
the best set of parameters is: age of 2.3 Gyr, $E(B-V)=0.145\pm
0.025$, and distance modulus $(m-M)_{o}=14.6$.  A very similar result
can be achieved by using a metallicity $Z=0.02$, provided that all the
main parameters be readjusted: age 2.5 Gyr, differential reddening
$E(B-V)=0.025\pm 0.025$ and distance modulus $(m-M)_{o}=14.8$.

\begin{figure*}
    \includegraphics[width=15cm]{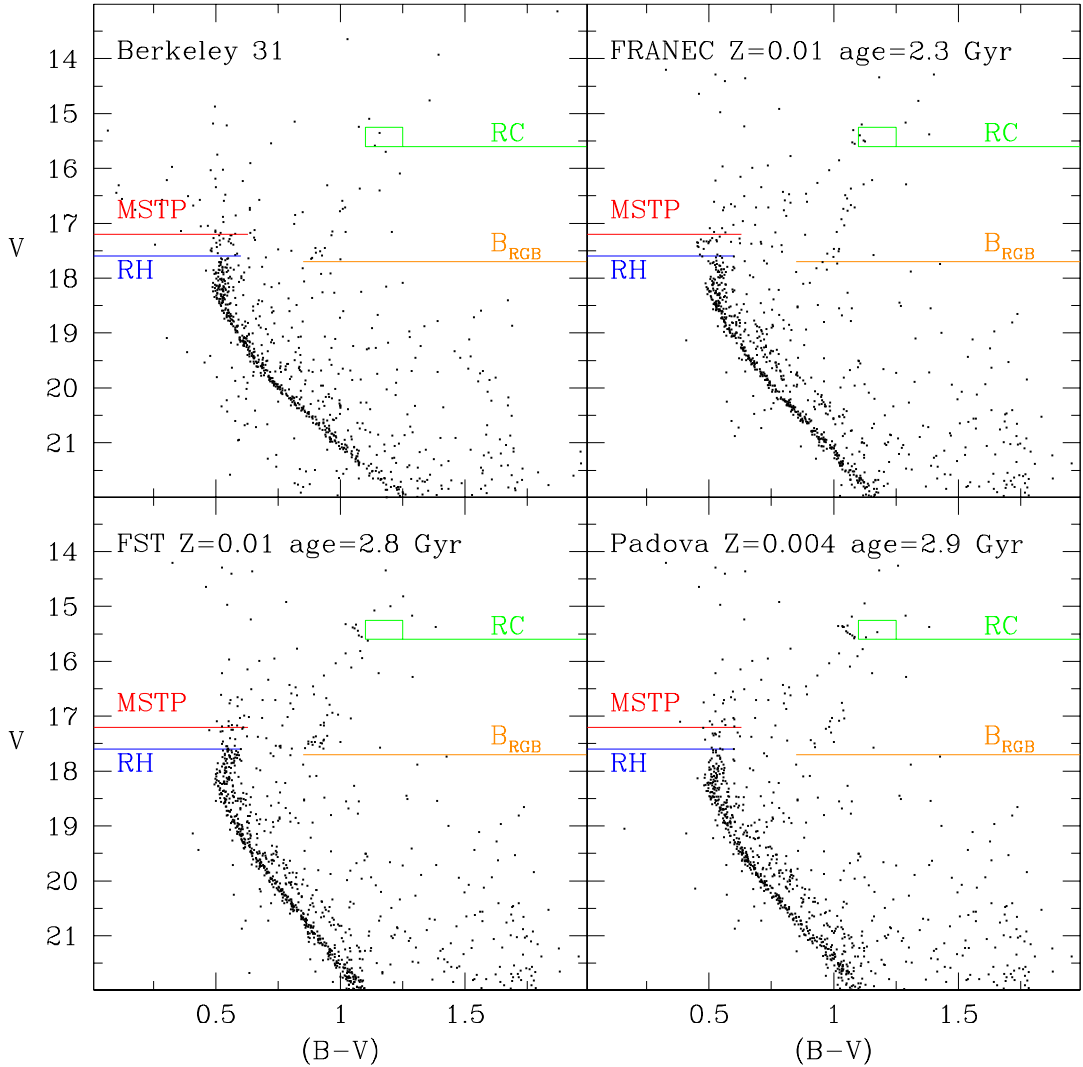} 
  \caption{Top left panel: CMD of stars within $2.5^{\prime}$ of the
    centre of Be~31. The other panels, clockwise from this, show the
    best fitting synthetic CMDs for the following parameters: FRANEC
    $Z=0.01$, age 2.3 Gyr, $E(B-V)=0.145\pm 0.025$ and $(m-M)_0=14.6$;
    Padova $Z=0.004$, age 2.9 Gyr, $E(B-V)=0.185 \pm 0.025$ and
    $(m-M)_0=14.4$; FST $Z=0.01$, age 2.8 Gyr, $E(B-V)=0.095\pm 0.025$
    and $(m-M)_0=14.6$. The adopted percentage of binaries (with
    random mass ratio) is always 24\%.}
  \label{be31_data_model_2p5}
\end{figure*}

Figure \ref{be31_data_model_2p5} summarises the best fitting CMD for
each set of tracks, compared with the observational CMD (top left
panel).

Trying to understand the results in terms of different prescriptions
of the input physics, a first interesting conclusion concerns the
efficiency of the core overshooting. As expected, the RH length is
severely affected by this macroscopic effect as a consequence of the
longer core hydrogen burning. Models with overshooting predict ages
higher (2.6-2.9 Gyr) than those inferred with canonical models
(2.3-2.5 Gyr). Our analysis suggests that the observed RH morphology
is better delineated by the non-overshoot tracks, as represented by
the FRANEC models, while the Padova and FST solutions, which use
different amounts of overshooting, are too hooked. On the other hand,
the low MS is unaffected by overshooting, being these stars
characterised by radiative cores. In this case, the good match offered
by low mass FRANEC models reflects, rather, a combined effect of the
equation of state and of atmosphere model.

 On the basis of this discussion, the FRANEC set should be preferred,
 since it is in agreement with both the RH morphology and with the low
 MS location. This restricts the set of possible ages to the range
 from 2.3 to 2.5 Gyr. Concerning the metallicity, a solar abundance is
 ruled out by a clear inability to reproduce the RGB colour. This
 result is in reasonable agreement with what has been found in
 literature, both by photometric and spectroscopic methods (see
 Sect. 1).

With these ranges of age and metallicity, the mean Galactic reddening
is between 0.145 and 0.175 mag, only slightly higher than the
\cite{sfd98} estimate $E(B-V)=0.145$ (which is an upper limit, being
the asymptotic reddening in the cluster direction, but which is also
more uncertain at low Galactic latitudes like those of our clusters),
while the best estimate for the distance modulus is between 14.46 and
14.6.

 A final comment concerns the choice of adopting a fraction of
 binaries of 24\%: surprisingly, a visual inspection of Fig. \ref{be31_data_model_2p5} seems to suggest that this number is
 slightly too high. However, this is clearly an artifact of the
 minimum mass ($0.6\,M_{\odot}$) in our stellar models, whose effect
 is to bias the binary distribution toward the high mass
 ratio. Although this effect is negligible near the TO, it grows near
 the lower end of the MS.

\subsection{Berkeley 23}

\begin{figure*}
    \includegraphics[width=14cm]{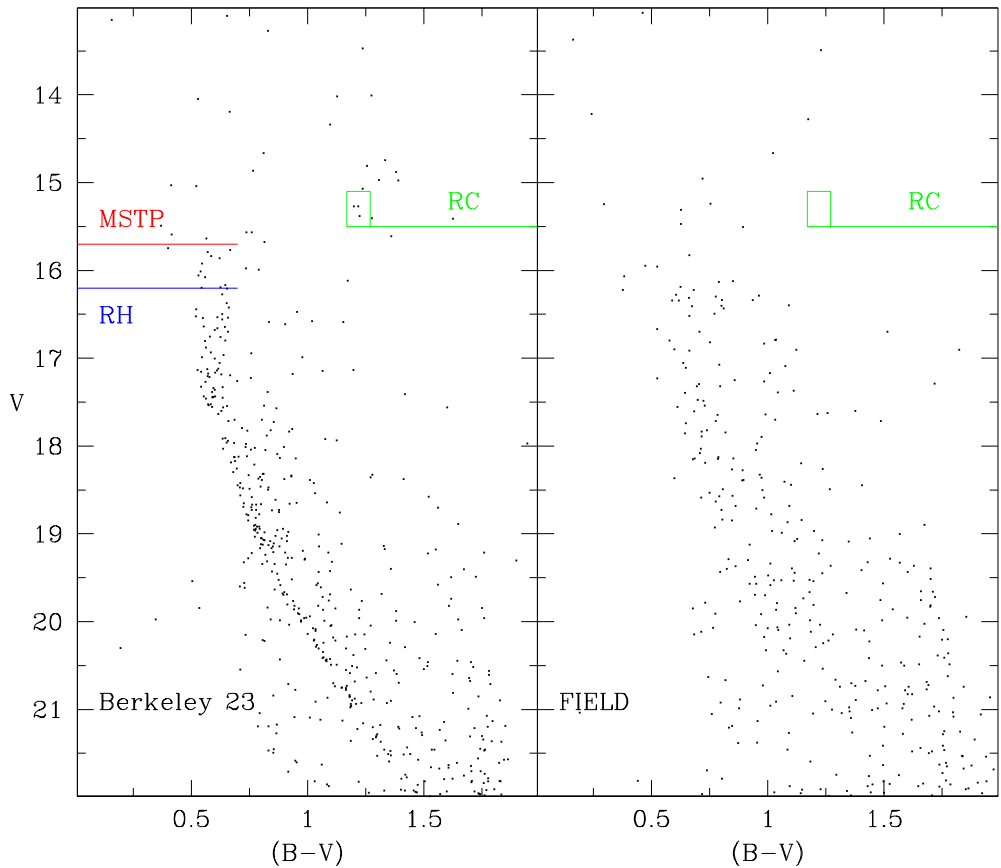} 
  \caption{Left panel: CMD of stars within $3^{\prime}$ of the centre
    of Be~23. Labels indicate the luminosity level of the RC, the MSTP
    and the RH. Right panel: control field CMD (taken from an equal
    area adjacent region).}
  \label{be23_3}
\end{figure*}

The strong contamination and the apparent small number of cluster
stars make the CMD of Be~23 rather poorly defined. In the left panel
of Fig. \ref{be23_3} we show the CMD restricted to the inner 3\arcmin
\ from the cluster centre (that is about three $r_c$ and slightly less
than $r_h$, see Fig.~\ref{profile} and Table~\ref{tab_param}).  This
partially cleans the diagram, revealing some signatures of the
cluster: 1) a broad and irregularly clumped MS, clearly
distinguishable at least down to $V=22$; 2) a RH positioned near
$V=16.2$ (evidenced in blue); 3) a MSTP near $V=15.7$ (in red); 4) a
mild excess of stars which is compatible with an apparently elongated
RC, extending between $V=15.1$ and $V=15.5$ (green). The RGB, not
sufficiently populated to be distinguishable from the field, is
instead difficult to recognise. This circumstance, together with the
wide OvC gap and the extended RC are indications that the cluster is
younger than 2 Gyr.

Before applying the synthetic CMD approach, we have evaluated the
amount of differential reddening and the fraction of binaries. The MS
presents a substantial intrinsic scatter, larger than expected from
photometric errors, suggesting a differential reddening of at least
0.05 mag. The percentage of binaries is not easy to obtain, given the
strong field contamination. Using the same strategy applied to Be~31,
we have found a range of values between 20\% and 30\%. Such large
uncertainty should be taken as a consequence of the spatial
fluctuations across the control field. In the following we adopt a
differential reddening $E(B-V)=0.05$ and a mean fraction of binaries
of 25\%.

In order to put limits on the cluster age and metallicity, the
3\arcmin \ CMD is compared with our synthetic CMDs.  Generally
speaking, all examined models do not satisfactorily reproduce two
evident CMD features. First of all, the observed RH morphology is more
vertical than predicted by the models. Furthermore, we cannot effectively
reproduce the number counts at the MSTP and at the RC, which are
always under-predicted by all models.

Despite these discrepancies, some conclusions can be drawn using the RH,
MSTP and RC luminosities as well as the MS colour. Table \ref{t2}
summarises the best combination of parameters and the residual
discrepancies for each set of tracks.

\begin{table*}
 \centering
  \caption{Be~23 best fit parameters (see previous table).}
\setlength{\tabcolsep}{1.5mm}
  \begin{tabular}{@{}lllllllll@{}}
  \hline
   Set&Z&Age (Gyr) & $E(B-V)_{\mathrm{M}}$ & $E(B-V)_{\mathrm{D}}$ &$(m-M)_0$& RH(C)&RC (C)&Low MS (C)\\
 \hline
 FST ($\eta=0.2$)&0.02  &1.3   &0.225 &$\pm 0.025$    &14.00  &OK&OK &TB (V$>$20) \\
 FST ($\eta=0.2$)&0.01  &1.2   &0.325 &$\pm 0.025$    &13.80  &TB&OK &TB (V$>$20)\\
 FST ($\eta=0.2$)&0.006 &1.1   &0.385 &$\pm 0.025$    &13.75  &TB&OK &TB (V$>$20)\\

 Padova&0.02  &1.3   &0.225 &$\pm 0.025$   &14.00 &OK&TR &TB (V$>$20)\\
 Padova&0.008 &1.3   &0.325 &$\pm 0.025$   &13.75 &TB&OK &TB (V$>$20)\\
 Padova&0.004 &1.2   &0.425 &$\pm 0.025$   &13.6  &TB&OK &TB (V$>$20)\\

 FRANEC&0.02  &1.0   &0.245&$\pm 0.025$   &13.95 &TB&OK &TB (V$>$21)\\
 FRANEC&0.01  &0.9   &0.375&$\pm 0.025$   &13.75 &TB&OK &TB (V$>$21)\\
 FRANEC&0.006 &0.8   &0.445&$\pm 0.025$   &13.75 &TB&OK &TB (V$>$20.5) \\
\hline
\end{tabular}
\label{t2}
\end{table*}

All the examined models can be reconciled with the luminosity
constraints, whereas only the FST and the Padova tracks at solar
metallicity can fit the average RH colour and the RC colour. In both
cases the cluster age is found to be around 1.3 Gyr. At lower
metallicities the synthetic RH colours become systematically too
blue. This drawback is particularly severe for the FRANEC tracks. The
reason for this is the younger ages required for the FRANEC set to fit
the luminosity constraints (RH, MSTP, RC), since they do not consider
overshooting. However, the FRANEC models work better than any other to
reproduce the colour of the lower MS. In particular, these tracks are
able to match the MS down to $V\approx\,21$ while all other models
diverge to the blue below $V\approx\,20$.

\begin{figure*}
    \includegraphics[width=15cm]{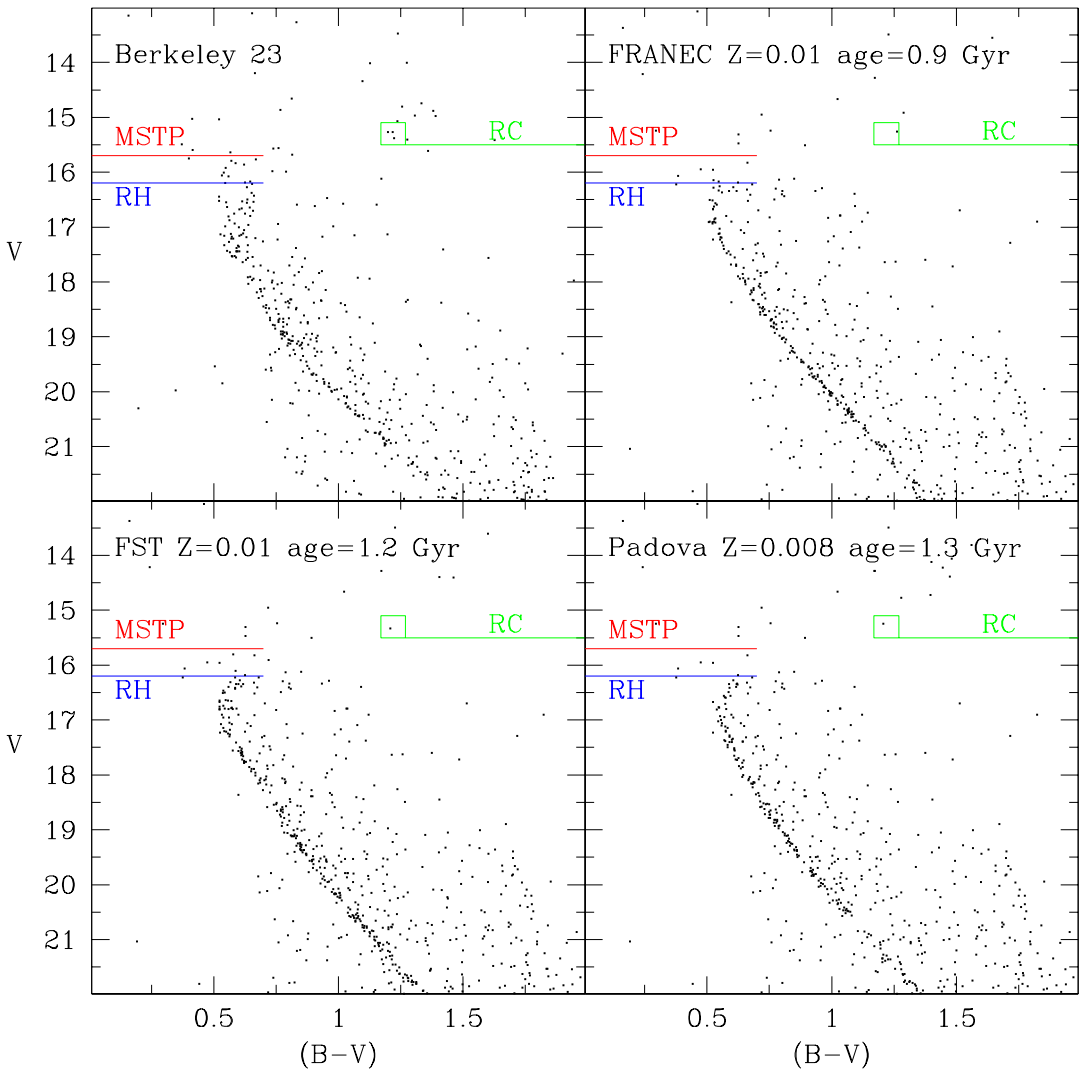} 
  \caption{Top left panel: CMD of stars within $3^{\prime}$ of the
    centre of Be~23. The other panels, clockwise from this, show the
    best fitting synthetic CMDs for the following parameters: FRANEC
    $Z=0.01$, age 0.9 Gyr, $E(B-V)=0.375\pm 0.025$ and
    $(m-M)_0=13.75$; Padova $Z=0.008$, age 1.3 Gyr, $E(B-V)=0.325\pm
    0.025$ and $(m-M)_0=13.75$; FST $Z=0.01$, age 1.2 Gyr,
    $E(B-V)=0.325\pm 0.025$ and $(m-M)_0=13.8$. The adopted percentage
    of binaries (with random mass ratio) is always 25\%.}
  \label{be23_best}
\end{figure*}

Figure \ref{be23_best} shows the best fitting CMD for each set of
tracks and the corresponding parameters.
We have decided to give preference to the RH region, which
provides information on the age, rather than the low MS fit, hence all
FRANEC models have been rejected. With this assumption, the age for
Berkeley 23 is well constrained between 1.1 and 1.3 Gyr, while the
favoured metallicity is solar. Concerning reddening and distance, the
range of variation for the former is 0.225-0.425 and for the latter is
between 13.6 and 14.00. Furthermore, it is noteworthy that the
reddening required for the solar models ($E(B-V)=0.225\pm 0.025$) is
too low compared with the \cite{sfd98} value ($E(B-V)=0.35$), while
that implied by the half solar models is perfectly consistent, and the
reddening required by even metal poorer models (Z=0.006 and Z=0.004)
is barely acceptable or too high. Combining these arguments we suggest
that the metallicity of Be~23 is probably between solar (as suggested
by the CMD fitting) and half solar.

As already noted in Sect. 1, there is discrepancy between different
analyses of this cluster, and our study makes no exception. We
disagree with \cite{ann2002}, who find the cluster to be much younger
and metal-richer, yet more reddened, even if these authors used MSTP
and RH luminosities very close to the values adopted here (see their
Fig.~3). However, their best fit model seems to miss the RC
position. We disagree also with \cite{hasegawa2004} on the age and
metallicity (their best solution is much older and metal-poorer) even
if reddening and distance modulus are in very good agreement.  The age
and metallicity differences arise mainly from a different
interpretation of the MSTP position: by looking at their Figure 3 it
is evident that these authors have identified as the MSTP a point that
is half a magnitude fainter than ours.


\subsection{King 8}

\begin{figure*}
    \includegraphics[width=14cm]{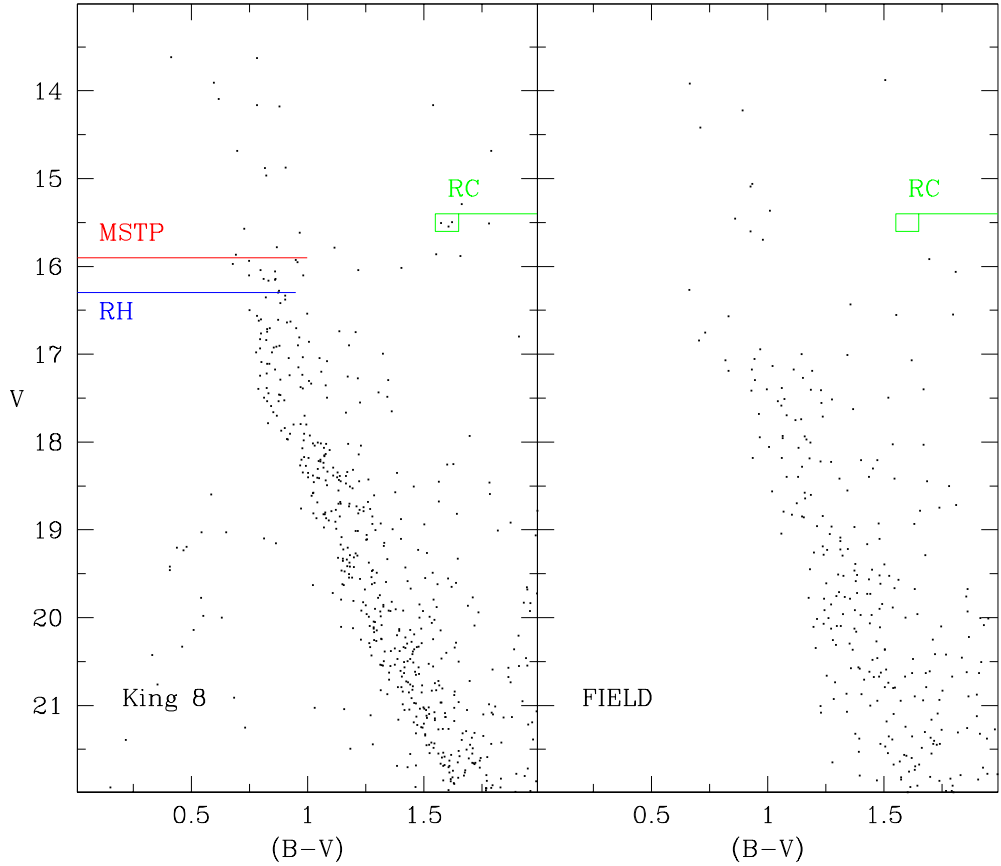} 
  \caption{Left panel: CMD of stars within $3^{\prime}$ of the centre
    of King~8. Right panel: control field CMD (as evaluated from an
    equal area adjacent region).}
  \label{king8_3m}
\end{figure*}

King~8 exhibits a very scattered CMD suggesting the strongest
differential reddening among the three clusters. This is clearly
evident from the left panel of Fig. \ref{king8_3m} where the CMD of
stars within 3\arcmin \ from the cluster centre (that is, three times
$r_c$, 2.5 $r_h$ and within $r_t$, see Fig.~\ref{profile} and Table 3)
is compared with the control field CMD (right panel).
In these conditions few CMD features are available for the fit: 1) a
MSTP around $V=16.0$ (indicated with a red line in Fig. \ref{king8_3m}); 2) a wide OvC gap; 3) a RC around $V=15.5$ (indicated
with a green box). As for Be~23, there is no evidence of a well
formed RGB. More importantly, the large MS spread prevents a safe
identification of the RH, which is tentatively identified as the mildly
hooked feature around $V=16.3$ (indicated with a blue line).

Given the wide (0.2 mag) and rather uniformly filled MS, the current
data do not allow us to reach a firm conclusion about the binary
fraction. Hence, in the following simulations we have assumed a fixed
fraction of 30\% (with random mass ratio). On the other hand, a
differential reddening $E(B-V)$ of at least 0.2 mag represents the
only viable solution to reproduce the observed MS spread. Keeping
fixed these parameters, we have investigated the possibility to fit
simultaneously the MSTP, the RH and the RC luminosities by adjusting
the age, the mean Galactic reddening and the distance modulus. The
best fitting results for the various sets and metallicities are
displayed in Table \ref{t3}, with the last two columns describing how
close the models match the RC and the lower MS colours.
\begin{table*}
 \centering
  \caption{King~8 best fit parameters. Columns 7 and 8 indicate the
    ability of the models to reproduce the RH colour and the lower MS
    colour. Additionally, the 8th column reports the magnitude below
    which the synthetic MS diverges.}  \setlength{\tabcolsep}{1.5mm}
  \begin{tabular}{@{}lllllllll@{}}
  \hline
   Set&Z&Age (Gyr) & $E(B-V)_{\mathrm{M}}$ & $E(B-V)_{\mathrm{D}}$ &$(m-M)_0$&RC (C)&Low MS (C)\\
 \hline
 FST ($\eta=0.2$)&0.02  &1.2   &0.55 &$\pm 0.10$    &13.20  &TB &TB (V$>$21) \\
 FST ($\eta=0.2$)&0.01  &1.2   &0.65 &$\pm 0.10$    &13.00  &TB &TB (V$>$21)\\
 FST ($\eta=0.2$)&0.006 &1.1   &0.75 &$\pm 0.10$    &12.95  &OK &TB (V$>$21)\\

 Padova&0.02  &1.3   &0.52 &$\pm 0.10$   &13.20 &OK &TB (V$>$21)\\
 Padova&0.008 &1.2   &0.65 &$\pm 0.10$   &12.90 &TB &TB (V$>$21)\\
 Padova&0.004 &1.1   &0.75 &$\pm 0.10$   &12.80 &TB &TB (V$>$21)\\

 FRANEC&0.02  &0.9   &0.60&$\pm 0.10$   &13.20 &OK &TB (V$>$22)\\
 FRANEC&0.01  &0.9   &0.67&$\pm 0.10$   &13.00 &OK &TB (V$>$22)\\
 FRANEC&0.006 &0.8   &0.78&$\pm 0.10$   &12.85 &OK &TB (V$>$22) \\
\hline
\end{tabular}
\label{t3}
\end{table*}

\begin{figure*}
    \includegraphics[width=15cm]{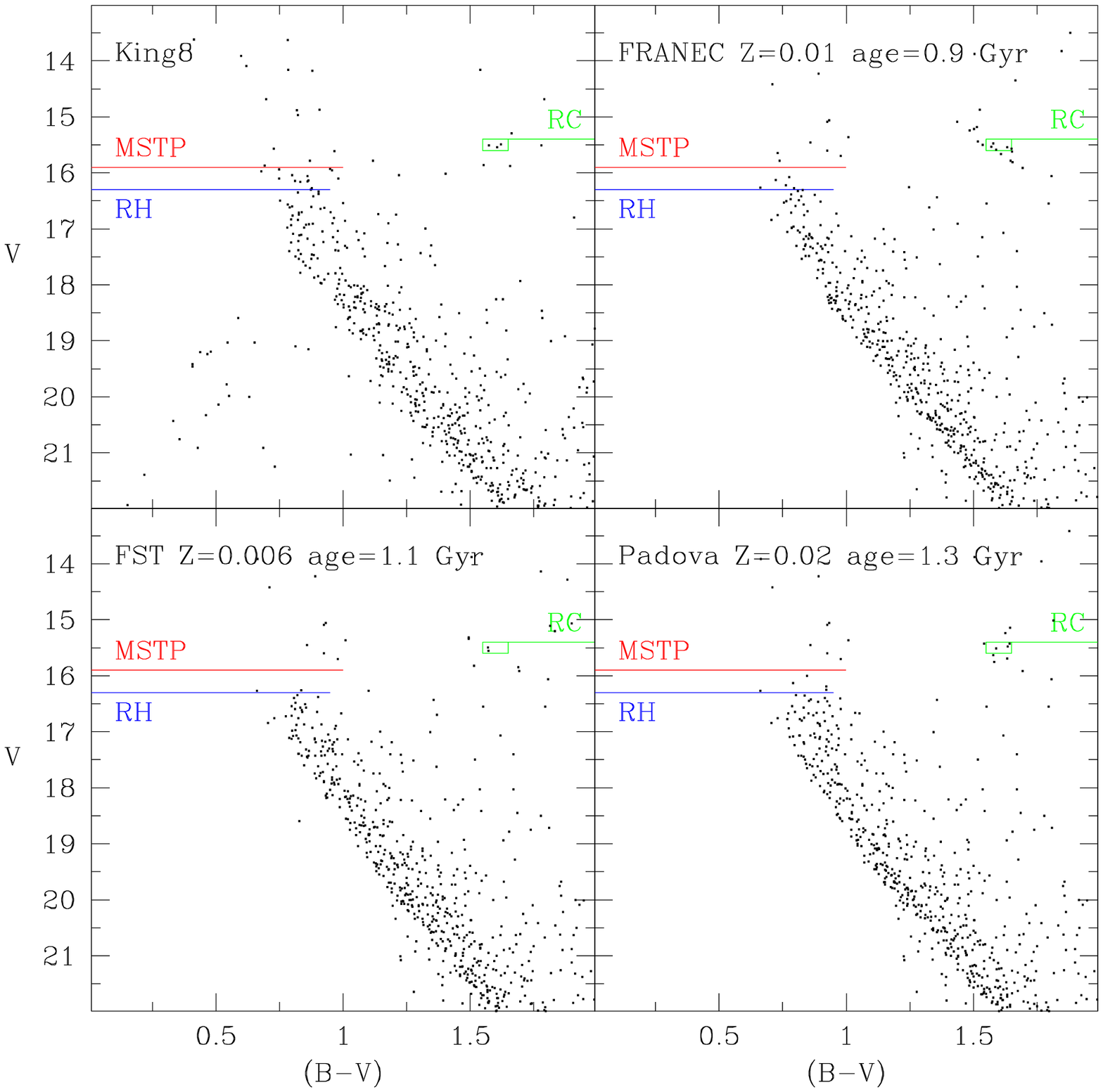} 
  \caption{Top left panel: CMD of stars within $3^{\prime}$ of the
    centre of King~8. The other panels, clockwise from this, show the
    best fitting synthetic CMDs for the following parameters: FRANEC
    $Z=0.01$, age 0.9 Gyr, $E(B-V)=0.67\pm 0.10$ and $(m-M)_0=13.0$;
    Padova $Z=0.02$, age 1.3 Gyr, $E(B-V)=0.52\pm 0.10$ and
    $(m-M)_0=13.2$; FST $Z=0.006$, age 1.1 Gyr, $E(B-V)=0.75\pm 0.10$
    and $(m-M)_0=12.95$. The adopted percentage of binaries (with
    random mass ratio) is always 30\%.}
  \label{king8_b}
\end{figure*}

In contrast with what is found for Be~23 and Be~31, all explored
models match very well the lower MS (at least up to $V=21$). On the
other hand, as a consequence of the high differential reddening, all
synthetic CMDs show a rather elongated RC while the observed RC is
round and concentrated. Fig. \ref{king8_b} summarises the best
fitting CMD for each set of tracks compared with the data CMD (top
left panel).

With the FST models, both the solar and the half-solar metallicities
fail to reproduce the RC colour (bluer than observed) and the number
of stars around the MSTP phase (systematically underestimated). A much
better result is achieved using the $Z=0.006$ models, whose synthetic
CMDs account for the RC colour, although still predicting too few MSTP
stars. In terms of age, the $Z=0.006$ case corresponds to about 1.1
Gyr, while the best fit distance is about
$(m-M)_0=12.95$. Furthermore, the high reddening implied by the
metallicity $Z=0.006$ is in good agreement with the literature value
($E(B-V)=0.8$) based on the \cite{sfd98} maps, while the reddening
solutions $E(B-V)=0.55\pm 0.10$ and $E(B-V)=0.65\pm 0.10$, required by
the metal richer tracks, are too low or barely acceptable, respectively.

In contrast with the FST result, among the Padova models only those at
solar metallicity predict the correct RC colour, whereas the synthetic
RC is too blue (by about 0.1 mag) at $Z=0.008$ and $Z=0.004$. On the
other hand, no significant difference is found in the best fit age
with varying metallicity: with $Z=0.02$ the best CMD is for 1.3 Gyr,
while with $Z=0.004$ it is 1.1 Gyr, results which are almost identical
to the FST findings. Focusing on the MSTP, both Padova and FST tracks
share the same mismatch (which is to be ascribed to the overshooting):
the predicted number of MSTP stars is always underestimated. As found
for the FST models, the comparison with the \cite{sfd98} maps
reddening favours the lowest metallicity ($Z=0.006$), in contrast
with our photometric estimate.

As far as the FRANEC models are concerned, we don't find a clear
indication to prefer one metallicity over the others. Both the
observed MSTP counts and the lower MS colours are fairly well
reproduced (the latter down to $V\approx\,22$), while the mismatch
here concerns the predicted number of RC stars (whose colours are in
good agreement with the observations), which outnumber the observed
counts by a factor of 3. The resulting ages (0.8 - 0.9 Gyr) are
younger than those obtained with the Padova and FST models, as a
direct consequence of the lack of overshooting. Once again, only the
sub-solar models ($Z=0.01$ and $Z=0.006$) imply a reddening value
compatible with the \cite{sfd98} estimate.

Also for this cluster the comparison to previous results is not
simple.  We seem to agree, apart from the Padova solution, on a rather
low metal content. The apparent agreement with
\cite{christian1981,christian1984} is not significant, given the
differences between the two photometries (see
Fig. \ref{confking8}). \cite{koposov} adopted solar isochrones for all
their clusters and this could at least partly explain the differences,
although the dependence on metallicity should be smaller in the near
IR bands.


\section{Conclusions}

\subsection{Galactic considerations}

Be~31 and Be~23 are rather far away from both the Galactic centre and
the Galactic plane. Berkeley 31 has a Galactocentric
distance\footnote{Assuming a Galactocentric distance of the Sun of 8
  kpc.} ($R_{GC}$) between 15.4 kpc and 15.9 kpc and a height above
the Galactic plane between 693 pc and 739 pc, while Berkeley 23 has a
$R_{GC}$ between 13.1 kpc and 14.2 kpc and a height between 494 pc and
594 pc. Such scale heights could be the result of a formation in situ
as a part of the Thick Disc population, as also suggested by the old
age of Be~31 (well over 2 Gyr), or due to a disc flaring which
gradually increases the Galactic scale-height of the Thin Disc. On the
other hand, the shorter scale height (between 196 pc and 236 pc), its
$R_{GC}$ between 11.6 kpc and 12.3 kpc and the younger (probably less
than 1 Gyr) age make King~8 a likely member of the Thin Disc.

Such findings help to interpret the difference in the field population
between the direction of Be~23 and Be~31 and the direction of King~8
visible in the lower panels of Fig. \ref{cmdall}. In the former, a
structure to the blue of the clusters' main sequence is clearly
visible starting from $V = 19$. Such a contamination is a signature of
a much farther and probably older population along the line of
sight. The most likely candidate is the above mentioned Thick Disc
(see e.g. \citealt{cignoni2008}), whose scale height is at least three
times longer than the corresponding Thin Disc. Although less probable,
it may also represents a tidal debris of a disrupted galaxy (see
e.g. the Monoceros ring, \citealt{Martin2004}) or a signature of the
Galactic Warp and Flare \citep{Momany2006}. On the other hand, the
lack of this feature in the direction of King~8 is mostly due to the
closer proximity to the Galactic plane (3 degrees against $\approx 5$
degrees of Be~23 and Be~31). At this latitude the line of sight goes
deeper into the Thin Disc (for about 6 kpc) before crossing a pure
Thick Disc sample, making the Thin Disc stars the dominant
contaminant. Furthermore, the Galactic reddening increases with
distance; this explains the elongated shape of the M-dwarfs sequence
(around $B-V\sim 1.9$) exhibited by the King~8 CMD field, a
circumstance that is not observed either in Be~23 or in Be~31 (where
the M-dwarf sequence is rather vertical).

\subsection{Summary}
Our study of open clusters is aimed at better understanding the
chemical and structural properties of the Galaxy. In this paper we
have studied three distant open clusters toward the Galactic
anti-centre using deep LBT photometry. The CMDs resulting from these
data are more precise and more than three mag deeper than the ones found
in literature, extending to more than six mag below the MS TOs. The
synthetic CMD technique has been used to derive a confidence interval
for age, metallicity, binary fraction, reddening, and distance for
each cluster. To remove the model dependence of the results three
different sets of stellar tracks (Padova, FST, FRANEC) have been
adopted. From the comparison between models and data we have drawn the
following conclusions:
\begin{itemize}

\item Be 31 is located at about 15-16 kpc from the Galactic
  centre and about 700 pc above the Galactic plane. The resulting age
  varies from 2.3 Gyr to 2.9 Gyr, depending on the adopted stellar
  model, with better fits for ages between 2.3 Gyr and 2.5
  Gyr. Concerning the metallicity we obtained a good match using
  the sets of models with metal content lower than solar. The mean
  Galactic reddening $E(B-V)$ is between 0.095 and 0.185 mag and the
  fraction of binaries is between 22\% and 26\%.

\item Be 23 is at about 13-14 kpc from the Galactic centre and 500-600
  pc above the plane. The age is well constrained between 1.1 and 1.3
  Gyr, while the best fitting metallicity is more uncertain: the CMD
  fit suggests a solar value, while only a half-solar metallicity is
  compatible with the \cite{sfd98} estimate for the Be 23's
  direction. The mean Galactic reddening $E(B-V)$ is between 0.225 and
  0.425 and the binary fraction is between 20\% and 30\%.

\item King 8 is at about 12 kpc from the Galactic centre and about 200
  pc above the plane. The strong differential reddening up to
  $\pm\,0.1$ mag broadens the colour extension of the MS, hindering a
  precise estimate of the cluster parameters. In this condition, the
  best fitting age is 0.8-1.3 Gyr, while the entire range of
  metallicities (0.004-0.02) is consistent with the data. On the other
  hand, only sub-solar models lead to reddening estimates (ranging
  from $E(B-V)=0.55$ to $E(B-V)=0.88$) in agreement with
  \cite{sfd98}. A 30\% of binaries is compatible with the observed
  CMD.

\end{itemize}

With these results, Be~31 and Be~23 are candidate Thick Disc open
clusters, while King~8 is a more classical Thin Disc member. Only
further direct spectroscopic and kinematic searches will allow to shed
light on the Galactic origin of these clusters.  They will also be
very important for any study of the radial distribution of metallicity
(and chemical abundances) in the Galactic disc, a crucial ingredient
in chemical evolution models.

Our choice to exploit the RC position even if it is barely visible in
the three clusters deserves a final comment. For Be~31, which shows
the least defined RC, many other relevant and well defined CMD
features (RH, MSTP, RGB, sharp MS) are available. Hence, even without
using the RC we would get the same parameter estimates. Moreover, the
\cite{sfd98} estimate of the reddening in this direction can be safely
used as an additional constraint. On the other hand, the CMDs of Be~23
and King~8 are both sufficiently contaminated and affected by
reddening that the RC position is a necessary constraint to derive
their ages. Relaxing the assumption on the RC position greatly expands
the range of possible ages and the only derivable result is that both
clusters are certainly younger than 2 Gyr. Still, we believe that,
although admittedly few, the stars in the CMD RC region are most
likely cluster members and can therefore be used in the analysis. A
definitive confirmation clearly needs proper motion or radial velocity
studies.

With the present study we may have reached the limit of what is
possible to derive simply with photometric data. Next step will be
provided by Gaia, the satellite due to fly in about two years, which
will produce photometry for about 10$^9$ Galactic objects, and most
importantly, parallaxes and proper motions of unprecedented
precision. The expected performances of Gaia will permit to obtain
individual distances of RC stars better than 10\% up to 8-10 kpc from
the Sun, i.e., including almost the entire family of known OCs.  The
precision on proper motions will be even better (the estimated error
on proper motions is about half the one on parallax), so that
membership for a large fraction of stars will be available also for
distant clusters. This will be important especially in cases like
Be~31, where the separation in radial velocity between field and
cluster stars is very small (see Introduction).

\section*{Acknowledgments}
We are grateful to the referee, Bruce Twarog, for his constructive and
helpful comments, that allowed to significantly improve our paper. We
thank Paolo Montegriffo for his software.  This paper has made use of
the WEBDA database, operated at the Institute for Astronomy of the
University of Vienna, of the SIMBAD database, operated at CDS,
Strasbourg, France, and of NASA's Astrophysics Data System. We are
grateful to the LBC team for the pre-reduction procedures.

\appendix

\end{document}